\documentclass[a4paper,12pt]{article}

\usepackage{citesort}

\usepackage{fleqn}
\usepackage{amsmath}

\def \lsim
{\raisebox{-3pt}{$\>\stackrel{<}{\scriptstyle\sim}\>$}}
\def \gsim
{\raisebox{-3pt}{$\>\stackrel{>}{\scriptstyle\sim}\>$}}

\def\beq{\begin{equation}}
\def\be{\begin{equation}}
\def\eeq{\end{equation}}
\def\ee{\end{equation}}
\def\bea{\begin{eqnarray}}
\def\eea{\end{eqnarray}}

\def\bq{\begin{quote}}
\def\eq{\end{quote}}
\def\nn{\nonumber}

\def\gappeq{\mathrel{\rlap {\raise.5ex\hbox{$>$}}
{\lower.5ex\hbox{$\sim$}}}}
\def\lappeq{\mathrel{\rlap{\raise.5ex\hbox{$<$}}
{\lower.5ex\hbox{$\sim$}}}}
\def\Toprel#1\over#2{\mathrel{\mathop{#2}\limits^{#1}}}

\allowdisplaybreaks

\begin{document}
\begin{titlepage}
\vspace*{-2.5cm}
\begin{flushright}
ROME1/1455-07
\\
DSF-NA/23/2007
\end{flushright}

{\Large
\begin{center}
{\bf Threshold Resummation in $B \to X_c \, l \, \nu_l$  Decays}
\end{center}
}
\vspace{.3cm}

\begin{center}
{U.~Aglietti~$^{a}$,~L.~Di~Giustino~$^{b}$,~G.~Ferrera~$^{c}$,
~A.~Renzaglia~$^{a}$,~G.~Ricciardi~$^{d}$,~L.~Trentadue~$^{b}$}
\\[7mm]
{$^{a}$\textit{Dipartimento di Fisica, Universit\`a di Roma I ``La Sapienza''
\\
and
\\
INFN Sezione di Roma, Roma, Italy }}
\\[4mm]
{$^{b}$\textit{Dipartimento di Fisica, Universit\`a di Parma,
\\
and
\\
INFN, Gruppo Collegato di Parma, Parma, Italy}}
\\[4mm]
{$^{c}$ \textit{  Departament ECM, Universitat de Barcelona, Barcelona, Spain
\\
and
\\
Dipartimento di Fisica, Universit\`a di Firenze, Firenze, Italy }}
\\[4mm]
{$^{d}$ \textit{Dipartimento di Scienze Fisiche, Universit\`a di
Napoli ``Federico II''
\\
and
\\ 
INFN, Sezione di Napoli, Napoli, Italy}} 
\\
[10pt] \vspace{0.5cm}
\begin{abstract}
We compute the QCD form factor resumming threshold logarithms in
$B \to X_c + l + \nu_l$ decays to next-to-leading logarithmic 
approximation.
We present an interpolation formula including soft as well as 
collinear effects softened by the non-vanishing charm mass which, in the $N$-space, reads:
\bea
\hskip -1.2truecm
\!\!\sigma_N\!
&=&\! \exp \int_0^1 dy \Big[ (1-y)^{N-1} - 1 \Big] 
\Bigg\{
\frac{1}{y} 
\int_{ \frac{Q^2 y^2}{1 + \rho} }^{ \frac{ Q^2 y^2 }{y + \rho} } 
A\left[ \rho ; \alpha_S(k^2) \right] \frac{d k^2}{k^2} + 
\frac{1}{y} D\left[ \alpha_S \left( \frac{Q^2 y^2}{1 + \rho} \right) \right] %+
\nn\\
&& ~~~~~~~~~~~~~
+ \, \left( \frac{1}{y} \, - \, \frac{1}{ y \, + \, \rho } \right)
\Delta \left[ \alpha_S \left( \frac{ Q^2 \, y^2 }{ y + \rho } \right) \right]  
 \, + \, 
\frac{1}{ y \, + \, \rho } \, B \left[ \alpha_S \left( \frac{ Q^2 \, y^2 }{ y + \rho } \right) \right]     
\Bigg\} \, ,
\nn\eea
where $y \, = \, \left( m_{X_c}^2 - m_c^2 \right)/\left( Q^2 \, - \, m_c^2 \right)$, 
$\rho \, = \, m_c^2/(Q^2 \, - \, m_c^2)$ and  
$Q \, = \, E_{X_c} \, + \, \left| \vec{p}_{X_c} \right|$.
The function $A(\rho; \alpha_S)$ describes soft radiation collinearly enhanced 
off the charm quark and reduces to the standard double-logarithmic function $A(\alpha_S)$ 
in the massless limit $\rho \to 0$.
The function $B(\alpha_S)$ describes hard collinear emission off the charm
quark, being the standard jet function.
$D(\alpha_S)$ and $\Delta(\alpha_S)$ describe soft radiation
not collinearly enhanced off the beauty and the charm quark respectively.
\end{abstract}
\end{center}
\end{titlepage}

\noindent

\section{Introduction and Summary of Results}
\label{sec1}

Distributions in semileptonic $B$ decays
\be
\label{sldecay}
B \, \to \, X_c \, + \, l \, + \, \nu_l \, , 
\ee
such as for example the hadron-mass event fraction, often
receive logarithmic contributions of infrared origin
of the form
\be
\label{generic}
\alpha_S^{\, n} \, \log^{\, k} \left( \frac{m_{X_c}^2 - m_c^2}{m_b^2 - m_c^2} \right) \, 
              \log^{\, l} \left( \frac{m_{X_c}^2}{m_b^2} \right) \, ,
\ee
where $m_{X_c}$ is the final hadron mass and $\alpha_S$ is the QCD coupling
evaluated at the hard scale.
Even in weak coupling regime $\alpha_S \ll 1 $ (perturbative phase), 
the terms above tend to spoil the convergence of the perturbative series in 
the threshold region, the latter being defined as the one having parametrically
\be
\label{thresh}
m_{X_c} \, \ll \, m_b \, .
\ee
The large infrared logarithms are a ``remnant'' of an incomplete cancellation
between real corrections to the decay distributions, integrated on the border
of the phase space, and virtual ones.
An all-order resummation is needed in order to have a reliable description
of the spectra in the whole kinematical range.
The infrared logarithms coming from soft-gluon emission are of the form
\be
\log \left( \frac{ m_{X_c}^2 - m_c^2 }{ m_b^2 - m_c^2 } \right)
\ee
and formally diverge for $m_{X_c} \to m_c^+$.
Since there is at most one soft logarithm for each gluon emission, i.e. for
each power of $\alpha_S$, it follows that $k \le n$ in eq.~(\ref{generic}).
The infrared logarithms of collinear origin,
\be
\log \left( \frac{ m_{X_c}^2 }{m_b^2} \right) \, ,
\ee
never become infinite but may become large in the threshold region (\ref{thresh}).
Since there are at most two infrared logarithms for each power 
of $\alpha_S$, $k + l \le 2 \, n$ in eq.~(\ref{generic}).

Our main result is the following expression for the resummed QCD form
factor in $N$-space (also called moment or Mellin space):
\bea
\label{main}
\label{finale}
\hskip -1.05truecm
\sigma_N (\rho, Q^2)
\!&=&\! \exp \int_0^1 dy \Big[ (1-y)^{N-1} - 1 \Big] 
\Bigg\{
\frac{1}{y} \, 
\int_{ \frac{Q^2 \, y^2}{1 \, + \, \rho} }^{ \frac{ Q^2 \, y^2 }{y \, + \, \rho} } 
A\left[ \rho ; \alpha_S(k^2) \right] \frac{d k^2}{k^2} + 
\frac{1}{y} D\left[ \alpha_S \left( \frac{Q^2 \, y^2}{1+\rho} \right) \right] +
\nn\\
&& ~~~~~~~ 
+ \, \left( \frac{1}{y} \, - \, \frac{1}{ y \, + \, \rho } \right)
\Delta \left[ \alpha_S \left( \frac{ Q^2 \, y^2 }{ y + \rho } \right) \right]  
 \, + \, 
\frac{1}{ y \, + \, \rho } \, B \left[ \alpha_S \left( \frac{ Q^2 \, y^2 }{ y + \rho } \right) \right]     
\Bigg\} \, .
\eea
We have defined the hadron variable with unitary range
\be
y \, \equiv \, \frac{m_{X_c}^2 - m_c^2}{ Q^2 \, - \, m_c^2 },
\ee
which equals zero in the Born kinematics, and the mass-correction parameter
\be
\rho \, \equiv \, \frac{m_c^2}{ Q^2 \, - \, m_c^2},
\ee
where $Q$, the hard scale in the heavy flavor decay, is given by
\be
Q \, \equiv \, E_{X_c} \, + \, \left| \vec{p}_{X_c} \right| \, ,
\ee
with $E_{X_c}$ and $\vec{p}_{X_c}$ being the total energy and 3-momentum of the final
hadron state $X_c$.
The function $A(\rho; \alpha_S)$ has an expansion in power of $\alpha_S$,
\be
A\left( \rho \, ; \alpha_S \right) \, = \, \sum_{n=1}^{\infty} A^{(n)}(\rho) \, \alpha_S^n \, ,
\ee
and describes soft radiation collinearly enhanced off the charm quark; it reduces to the 
standard double-logarithmic function $A(\alpha_S)$ in the massless limit:
\be
A(\rho; \alpha_S) \, \to \, A(\alpha_S) ~~~~~~ {\rm for} ~~~ \rho \, \to \, 0^+ \, .
\ee
The first-order coefficient is:
\be
A^{(1)}(\rho) \, \equiv \, \frac{C_F}{\pi} \, \left( 1 + 2 \rho \right) \, ,
\ee
where $C_F = (N_c^2-1)/(2N_c) = 4/3$ for $N_c=3$.
The second-order coefficient $A^{(2)}(\rho)$ (as well as the third-order one $A^{(3)}(\rho)$) 
is only known in the massless limit (see later).
The function $B(\alpha_S)$ is the standard massless jet-function, describing 
hard collinear emission off the charm quark, while $D(\alpha_S)$ and $\Delta(\alpha_S)$ 
describe soft radiation not collinearly enhanced off the beauty and charm quarks 
respectively. All these functions have a perturbative expansion in $\alpha_S$\footnote{
For a compilation of the massless coefficients, see for example 
\cite{Moch:2005ky}, while for the heavy flavor case \cite{Aglietti:2005mb}.
}:
\be
B(\alpha_S) \, = \, \sum_{n=1}^{\infty} B^{(n)} \, \alpha_S^n \, ;
~~~~~
D\left(\alpha_S \right) \, = \, \sum_{n=1}^{\infty} D^{(n)} \, \alpha_S^n \, ;
~~~~~
\Delta\left(\alpha_S \right) \, = \, \sum_{n=1}^{\infty} \Delta^{(n)} \, \alpha_S^n \, .
\ee
The $\mathcal{O}(\alpha_S)$ coefficients read:
\be
B^{(1)} \, = \, - \, \frac{3}{4} \, \frac{C_F}{\pi} \, ;
~~~~~~~
D^{(1)} \, = \, - \, \frac{C_F}{\pi} \, ;
~~~~~~~
\Delta^{(1)} \, = \, - \, \frac{C_F}{\pi} \, .
\ee
The form factor (\ref{main}) aims at describing the three different dynamical regions 
in the semileptonic decay (\ref{sldecay}), which are identified by the value of the 
(ordinary) charm velocity $u_c = p_c / E_c$ --- without any generality loss, we can work 
in the beauty rest frame. It holds:
\be
\rho \, \simeq \, \frac{1 - u_c}{2 u_c} \, .
\ee
These regions are:
\begin{enumerate}
\item
{\it very slow charm quark:}
\be
u_c \, \gsim \, 0 ~~~ {\rm or, ~ equivalently, } ~~~ \rho \, \gg \, 1 \, .
\ee
For $u_c \, = \, 0$ there is no {\it soft-gluon} emission to any order in perturbation
theory \cite{Isgur:1989vq}
\footnote{
By ``soft'' we mean a gluon with energy $E_g < m_c, m_b$.
}. 
That is a coherence effect, namely destructive interference between
emission off the initial and the final state.
The physical picture is the following.
The $b$ decays into the $c$ at rest and with 
the same color state of the $b$. 
Soft gluons ``view'' the $b$ as well as the $c$ as static color charges,
and therefore do not see any acceleration or color-spin flip, i.e. any change  
occurring in the decay. As a consequence,  there is no radiation.
Furthermore, for small $u_c$, radiation is proportional to $u_c^2$,
i.e. first order terms $\mathcal{O}(u_c)$ vanish 
\cite{Korchemsky:1987wg}.
Radiative corrections to our form factor vanish in the no-recoil point:
\be
\label{rto1}
\sigma_N(\rho; \, Q^2) \, \to \, 1 ~~~~~~ {\rm for} ~~~ \rho \, \to \, + \, \infty \, ,
\ee
implying that this region is correctly described by the form factor;
\item
{\it non-relativistic charm quark.} Schematically:
\be
u_c \, \approx \, \frac{1}{3} ~~~ {\rm or} ~~~ \rho \, \approx \, 1 \, .
\ee
The final state $X_c$ contains the final charm quark together with soft gluons 
emitted at any angle with respect to it, because there is a significant soft emission.
Since the QCD matrix elements do not have any collinear enhancement,
$X_c$ does not have a jet structure.
In the threshold region $y \ll 1$, so that $y \ll \rho$ 
and the form factor (\ref{main}) simplifies into:  
\bea
\label{simple}
\label{simplified}
\hskip -1.2truecm
\sigma_{S,\,N} (\rho,Q^2)
\!&=&\! \exp \!\int_0^1 \frac{dy}{y} \Big[ (1-y)^{N-1} - 1 \Big] 
\Bigg\{
\int_{ \frac{Q^2 \, y^2}{1 + \rho} }^{ \frac{ Q^2 \, y^2 }{ \rho } } 
\!A\left[ \rho ; \alpha_S(k^2) \right] \frac{d k^2}{k^2} +  
D\left[ \alpha_S \left( \frac{Q^2 y^2}{1 + \rho} \right) \right] %+
\nn\\
&& ~~~~~~~~~~~~~~~~~~~~~~~~~~~~~~~~~~~~~~~~~~~~~~~~~~~~~~~~~~~
+ \,
\Delta \left[ \alpha_S \left( \frac{ Q^2 \, y^2 }{ \rho } \right) \right]       
\Bigg\} \, ;
\eea
\item
{\it fast charm quark}:
\be
u_c \, \lsim \, 1 ~~~ {\rm or} ~~~ \rho \, \ll \, 1 \, .
\ee
Soft gluons are mostly radiated at small angle with respect to the charm
quark and there is a jet structure of the final state.
For $\rho \to 0^+$ the form factor (\ref{main}) reduces to the standard massless
expression \cite{Sterman:1986aj,catanitrentadue,Aglietti:2001br,Aglietti:2005mb}:
\bea
\label{vecchiamica}
\hskip -1truecm
\sigma_N (0, Q^2)
&=& \exp \int_0^1 \frac{dy}{y} \Big[ (1 - y)^{N - 1} - 1 \Big] 
\Bigg\{ 
\int_{ Q^2 \, y^2 }^{ Q^2 \, y } 
A\left[ \alpha_S(k^2) \right] \frac{d k^2}{k^2} \, + \,  
D\left[ \alpha_S \left(  Q^2 \, y^2  \right) \right] \, + \, 
\nn\\
%&& ~~~~~~~~~~~~~~~~~~~~~~~~~~~~~~~~~~~~~~~~~~~~~~~~~~~~~~~~~~ 
&\, + \,& B \left[ \alpha_S \left( Q^2 \, y  \right) \right]     
\Bigg\} \, ,
\eea
where:
\be
y \, \to \, u \, = \, \frac{E_X - \left| \vec{p}_{X} \right|}{E_X 
+ \left| \vec{p}_{X} \right|} \, \simeq \, \frac{m_X^2}{4E_X^2} \, ; 
~~~~~~~ Q \, \to \, E_X \, + \, \left| \vec{p}_{X} \right| \, \simeq \, 2 E_X \, ,
\ee
where on the last members we have expanded for $m_X \ll E_X$.
The inclusion of first-order corrections in $\rho$ on the r.h.s. 
of eq.~(\ref{vecchiamica}) basically amounts to
take into account the dead-cone effect in gluon radiation off the charm \cite{Aglietti:2006wh}.
The form factor factorizes into the product of the form factor for a massless
charm times the universal mass-jet correction:
\be
\sigma_N (\rho, \, Q^2) \, \simeq \, \sigma_N (0, \, Q^2) \, \delta_N(\rho, \, Q^2) 
~~~~~~~~~ {\rm for} ~~~ \rho \, \ll \, 1 \, ,
\ee
where
\footnote{
The variable $\rho$ used in this work and the variable $r \equiv m_c^2/Q^2 = \rho/(1+\rho)$ 
used in \cite{Aglietti:2006wh} coincide to first order.
}
\bea
\hskip -1truecm
\delta_N(\rho , Q^2) &=& \exp
\int_0^1 d y \frac{ (1-y)^{ \, \rho \, (N-1)} - 1 }{y}
\Bigg\{
 - \int_{ \rho Q^2 y^2 }^{ \rho Q^2 y } \frac{dk_{\perp}^2}{k_{\perp}^2} A\left[\alpha\left(k_{\perp}^2\right)\right]
 - B \left[\alpha\left( \rho Q^2 y \right)\right]
+ \nonumber\\
&&~~~~~~~~~~~~~~~~~~~~~~~~~~~~~~~~~~~~~~~~~~~~~~~~~~~~~~~~~~~
\, + \, D \left[\alpha\left( \rho Q^2 y^2 \right)\right]
\Bigg\}.
\eea
Let us note that in \cite{Aglietti:2006wh} the {\it universal} correction $\delta_N(\rho, Q^2)$
was found for $\rho \ll 1$, while in this work we consider a {\it specific} process for {\it any} 
$\rho$.
\end{enumerate}
Since the ratio $m_c/m_b \approx 1/3 \div 1/4$ in the real world, i.e. it is not so small, 
region 3. is tiny. With typical values of the on-shell masses, $m_b = 4.7$ GeV and 
$m_c = m_b - m_B + m_D \simeq 1.29$ GeV \cite{Aglietti:1991rr},
one has a maximal charm velocity $u_{c \,\max} \simeq 0.86$ corresponding to a 
Lorentz factor $\gamma_{c\,\max} \simeq 2$, or $0.081 \lsim \rho \le \infty$. 
With hadron kinematics, i.e. with $m_b = m_B$ and $m_c = m_D$, one obtains instead:
$u_{c\,\max} = 0.78$, $\gamma_{c\,\max} \simeq 1.6$ and $\rho_{\min} \simeq 0.14$.

\noindent
Let us comment further upon the properties of the form factor.

Since the $y$ and $k^2$ integrals on the r.h.s. of eq.~(\ref{main}) extend down to zero, 
one hits the infrared singularity in the QCD couplings $\alpha_S(k^2)$ and $\alpha_S(\cdots y^2)$ 
--- the well-known Landau pole:
some prescription is needed in order to render the integrands well behaved.
As it is usually the case with resummation formulae, 
eq.~(\ref{finale}), as it stands, has an ``algebraic'' sense:
upon expansion in powers of $\alpha_S$, it allows to predict the logarithmic corrections 
which would be explicitly found in higher-order Feynman diagram computations.

In general, we expect less radiation to be emitted in the decay (\ref{sldecay}), 
because of the rather large charm mass, compared to the charmless channel
\be
B \, \to \, X_u \, + \, l \, + \, \nu_l \, .
\ee
As a consequence, the typical Sudakov effects, namely suppression of non-radiative
channels and broadening of sharp structures, are expected to be less
pronounced for our process.
In principle, for the decay (\ref{sldecay}), one has single-logarithmic corrections,
which are not strong enough to shift the peak of tree-level distributions.
For $y\to 0^+$ they produce indeed a form factor of the form 
\be
\sigma(y) \, \approx \, \frac{d}{dy} e^{ - \alpha_S \, \log\frac{1}{y}}  
\, =  \, \frac{\alpha_S}{y^{ 1 - \alpha_S} } \, ,
\ee
which still has an (infinite) peak in $y=0$, like the lowest-order form factor $\sigma^{(0)}(y) =  \delta(y)$.
That is to be contrasted with a double-logarithmic form factor  
$\approx \, d/dy \exp ( - \alpha_S \, \log^2 y )$.  

It is remarkable that the limits of the $k^2$-integral on the r.h.s. of eq.~(\ref{main}),
$Q^2 y^2/(1+\rho)$ and $Q^2 y^2/(y+\rho)$, 
are equal to the arguments of the QCD couplings entering the ``sub-leading'' terms 
(i.e. those without the explicit $k^2$ integration); 
that seems a general feature of resummation formulae.
A consequence of this fact is that only 3 out of the 4 resummation
functions $A$, $B$, $D$ and $\Delta$ entering $\sigma_N$ are actually independent 
(see sec.~\ref{secinv}).
Similar considerations can be repeated for the soft form factor in eq.~(\ref{simplified}),
in which the scales $Q^2 y^2/(1+\rho)$ and $Q^2 y^2/\rho$ appear as limits of the
$k^2$ integral as well as arguments of $\alpha_S$ in the $D$ and $\Delta$ terms.
In the latter case, only 2 out of the 3 functions $A$, $D$ and $\Delta$ are independent.

By looking at the $y$-integration on the r.h.s. of eq.~(\ref{main}),  
one finds that the threshold region cannot be described within a ``pure'' perturbative framework
when the QCD coupling comes close to the infrared (Landau) singularity, namely when
\be
\frac{Q^2 \, y^2}{1 \, + \, \rho} \, \approx \, \Lambda_{QCD}^2 \, ,
\ee
where $\Lambda_{QCD}$ is the QCD scale, i.e. when the final hadron mass becomes as small as
\footnote{
For $Q = m_b$ and $\Lambda_{QCD} = 300$ MeV
one obtains for the current values of the $b$ and $c$ masses
$m_X \big|_{NP} \simeq 1.7$ GeV.}
\be
\label{slice}
m_{X_c}^2 \Big|_{NP} \, \approx \, m_c^2 \, + \, \Lambda_{QCD} \sqrt{Q^2 - m_c^2} \, .  
\ee
These effects are related to soft interactions only, because hard collinear terms
($\propto B(\alpha_S)$) are controlled by a smaller coupling, evaluated at the larger scale 
$Q^2 y^2/(y + \rho)$.
These soft interactions are non-perturbative in region (\ref{slice}), 
and can be factorized into the shape function for a massive final quark 
\cite{Mannel:1994pm}.
Note that for $m_c \to 0$, the slice (\ref{slice}) reduces to the 
well-known one $m_X^2 \, \approx \, \Lambda_{QCD} \, Q$ .
In region (\ref{slice}), also soft interactions between the $b$ quark and the light degrees
of freedom in the $B$-meson, namely the valence quark, are important.
That is the well-known ``Fermi-motion'', characterized by momentum exchanges $k_{\mu}$ 
in the $B$-meson of the order of the hadronic scale,
\be
| k_{\mu} | \, \approx \, \Lambda_{QCD} \, .
\ee
Physical intuition would suggest that Fermi motion is related to the initial $B$-meson
state only, i.e. that it is independent on the mass of the final quark.
In quantum field theory that is no true and
it is not easy to find a general relation between eq.~(\ref{vecchiamica}) 
with the $B$ term dropped, describing soft interactions for $m_c \to 0$, 
and eq.~(\ref{simplified}), describing soft interactions for $m_c \ne 0$.

It is technically simpler to compute first the form factor for the radiative case
\be
\label{radec}
B \, \to \, X_s \, + \, \gamma 
\ee
for $m_s \ne 0$ and then to extend the result to the semileptonic case (\ref{sldecay}), 
rather than to deal directly with the latter process. 
That is the strategy we will follow.

\noindent
The plan of this note is the following.

\noindent
In sec.~\ref{sec2} we consider  soft effects only, 
which are in some sense ``leading'', as they
become formally infinite at the border of the phase space, for 
the radiative decay (\ref{radec}).

\noindent
In sec.~\ref{sec3} we include  hard collinear effects,
which never become infinite but are logarithmically enhanced for small
strange mass, as discussed at the beginning of this section.

\noindent
In sec.~\ref{sec4} we derive the resummed form factor for the process of
main physical interest, namely the semileptonic decay (\ref{sldecay}), 
by means of a generalization of the radiative case.

\noindent
Finally, in sec.~\ref{sec5} we draw our conclusions and we give an outlook
to possible developments of the results, both of theoretical and phenomenological nature.

\section{Soft-Gluon Resummation}
\label{sec2}

In this section we consider the resummation of threshold logarithms
coming from multiple soft-gluon emission to all orders of perturbation 
theory.
The QCD form factor factorizing such terms has a perturbative expansion 
of the form:
\bea
\label{softff}
\sigma_S(y ; \, \rho , Q^2) 
&=& \delta(y) \, + \, \sum_{n=1}^{\infty} \sum_{k=0}^{n-1} c_{nk}(\rho) \, 
\alpha_S^{\, n}(Q^2) \left( \frac{\log^{\, k} y}{y} \right)_+
\nn\\
&=& \delta(y) \, + \, 
\frac{C_F \alpha_S}{\pi} \, \left[ \, \big(1 + 2\rho\big) \log \left( \frac{1+\rho}{\rho} \right) \, - \, 2 \right] 
\left(\frac{1}{y}\right)_+ \, + \, \mathcal{O}(\alpha_S^2) \, .
\eea
As discussed in the introduction, we first consider the simpler 
radiative decay (\ref{radec}), for which $y$ is naturally defined as
\begin{equation}
\label{y:def}
y \, \equiv \, \frac{m_{X_s}^2 - m_s^2}{m_b^2 - m_s^2}
\end{equation}
and the mass-correction parameter $\rho$ as
\be
\rho \, \equiv \, \frac{m_s^2}{m_b^2 - m_s^2} \, .
\ee
The $c_{nk}(\rho)$'s are functions of $\rho$ and the plus distributions are defined
as usual as:
\be
P(y)_+ \, \equiv \, \lim_{\epsilon \to 0^+} 
\Big[ \,  
\theta(y - \epsilon) P(y) - \delta(y - \epsilon) \int_{\epsilon}^1 dy' P(y')
\, \Big] \, .
\ee
Let us note that we do not have here a double-logarithmic problem 
as in the massless case ($\rho = 0$),
but a {\it single-logarithmic} one: there is at most one infrared logarithm
for each power of $\alpha_S$. 
The additional singularities for $\rho \to 0^+$ are ``hidden'' in the coefficients
$c_{nk}(\rho)$'s, which indeed diverge logarithmically in that limit --- see the last 
member in eq.~(\ref{softff}).
Note that the square bracket on the last member of eq.~(\ref{softff})
vanishes for $\rho\to\infty$ (see region 1. in the introduction).
By leading logarithmic soft ($LL_S$) approximation, we mean the resummation of all the
terms in eq.~(\ref{softff}) with $k = n-1$, by next-to-leading logarithmic ($NLL_S$) 
approximation also those with $k=n-2$ and so on.

\subsection{Single-Gluon Emission}

In this section we consider the emission of a real gluon, i.e. the process:
\be 
b \, \rightarrow \, s \, + \, g \, + \, \gamma \, . 
\ee 
We will recover the soft contributions factorized in $\sigma_S$ above.

\subsubsection{Kinematics}

The energy of the strange quark in the tree-level process
\be
\label{treelev}
b \, \rightarrow \, s \, + \, \gamma
\ee
is fixed to
\be
E_s^{(0)} \, = \, \frac{m_b}{2} \frac{1+2\rho}{1+\rho}
\ee
and its 3-momentum to
\be
p_s^{(0)} \, \equiv \, | \vec{p}_s^{\,\,(0)}  | \, = \, E_{\gamma}^{(0)} \, = \, \frac{m_b}{2 (1+\rho)} \, . 
\ee
The strange quark and the photon are emitted back to back in the decay of the $b$ 
quark at rest. Note that $E_s^{(0)} \, + \, p_s^{(0)} \, = \, m_b \, .$
Let us now consider kinematics of single-gluon emission.
The maximum value of the gluon energy is 
\be
E_g^{\max} \, = \, \frac{m_b}{2(1+\rho)}
\ee 
and it corresponds to the configuration with the strange quark
and the gluon back to back with an accompanying soft photon.
Let us define the unitary gluon energy 
$\omega$ as:
\be
\omega \, \equiv \, \frac{2 E_g}{m_b}  (1+\rho)
\ee
and the unitary angular variable $t$ as
\be
~~~~~~
t \, \equiv \, \frac{1 - \cos\theta}{2} \, \simeq \, \left( \frac{\theta}{2} \right)^2
~~~~~~~~~~~{\rm for}~\theta \, \ll \, 1 ,
\ee
where $\theta$ is the emission angle of the gluon with respect to
the final (on-shell) strange quark. 
Let us now restrict our attention to kinematics {\it in the soft limit}.
$E_s$ and $p_s$ are only mildly modified by the emission of a soft gluon, 
so that we can neglect recoil effects and set
\be
E_s \, \cong \, E_s^{(0)} \, ; 
~~~~~~~  
p_s \, \cong \, p_s^{(0)} .
\ee
Furthermore, the direction of motion of the strange quark is modified by gluon 
emission in negligible way.
In terms of the variables defined above, we can write 
\be
\label{expressy}
y \, = \, \omega\, \frac{t + \rho}{1+\rho} \, . 
\ee
Note the linear relation between $y$ and $\omega$, which is a consequence of the soft limit,
and that
\be
y \, \to \, 0 ~~~ \iff ~~~ \omega \, \to \, 0.
\ee
That implies that the threshold region $y \ll 1$ selects kinematical
configurations which are dynamically enhanced by the soft divergencies.
At fixed $y$, one has for the angular variable $t$ the whole range 
$0 \, < \, t \, < \, 1$.
Finally, note that for any hadron final state:
\be
\label{suggerimento}
E_{X_s} \, + \, \left| \vec{p}_{X_s} \right| \, = \, m_b \, .
\ee
That is a consequence of the massless ``probe'', namely the photon 
($q^2 = 0$ with $q = p_{\gamma}$).

\subsubsection{Eikonal Current}

In the soft limit $\omega \ll 1$, the emission
of a real gluon  off the $b$ and the $s$ quarks is described
by the eikonal current:
\be
J^{\mu} (k) \, = \, i g \, \left( {\bf T_b} \, \frac{p_b^\mu}{p_b \cdot k} \, - \, {\bf T_s} \,
\frac{p_s^\mu}{p_s \cdot k}\right) \, ,
\ee
where ${\bf T_b}$ and ${\bf T_s}$ are the color generators of the $b$ and $s$ quarks respectively
\cite{Aglietti:2006wh}.
By squaring the eikonal current, integrating the
phase space over the azimuth and summing/averaging over helicities and polarizations, 
we have the following contribution to the differential form factor from the emission of a real gluon:
\be
\label{real:single0}
\hskip -1truecm
\sigma^{(R)}_S(y)
 \, = \,  
\frac{C_F \, \alpha_S}{\pi}\,\int_0^1 dt \, \int_0^1 \frac{d\omega}{\omega} 
\left\{ 
 \frac{1 + 2\rho}{ t + \rho }
 - 1 - \frac{\rho \, (1 + \rho)}{ \left( t  + \rho \right)^2 }
\right\} \, \delta \left( y \, - \, \omega \frac{t + \rho}{1+\rho} \right) \, .
\ee
Let us briefly comment upon the result above.
The soft singularity for $\omega \to 0$ of the QCD matrix element squared
is screened by the kinematical
constraint, which is infrared safe ($y \ne 0 \Rightarrow \omega \ne 0$) \cite{qcdgeneral}.
Also the collinear singularity is screened by the kinematical constraint, which
is collinear safe ($y \ne 0 \Rightarrow t \ne 0$ for $\rho = 0$), as well as by the strange mass, 
because the expression (\ref{real:single0}) diverges for $t \rightarrow - \rho$
only, which is outside the physical domain ($t \ge 0$) for $\rho > 0$. 

\subsection{Inclusive Gluon-Decay Effects}
\label{igd}

In the previous section we considered the $\mathcal{O}(\alpha_S)$ correction to the tree-level
decay (\ref{treelev}), namely the emission of a single gluon.
Higher order corrections in $\alpha_S$ basically involve {\it multiple} emissions of the 
following two kinds:
\begin{enumerate}
\item
{\it primary} emissions, i.e. direct emissions of gluons directly attached to
the colored $b \to s$ line;
\item
{\it secondary} emissions, produced by the splitting of a gluon
into a $gg$, $q\bar{q}$ pair, or by gluon emission off secondary
quarks or antiquarks, etc.
\end{enumerate}
As we are going to explicitly show in sec.~(\ref{Nspace}), the main effect of
primary emissions is to exponentiate the single-gluon distribution.
Let us consider in this section secondary emissions, namely the ``decay'' of
a primary (virtual) gluon with momentum $p_g$ into secondary partons.

The amplitude for the emission of one gluon off the strange quark is enhanced when
the scalar part of the $s$-quark propagator,
\be
\Delta_s(p_s^*)\, = \, \frac{ 1 }{ { p_s^* }^2 - m_s^2},
\ee
becomes large, i.e. when the virtual $s$ quark is close to its mass-shell.
This denominator reads in the case of the emission of one virtual gluon with momentum 
$p_g$ and invariant mass squared $ m_g^2 = p_g^2$:
\bea
\label{varieapprox}
\Delta_s(p_s + p_g)
&=& 
\frac{1}{2 p_s \cdot p_g + m_g^2} 
\, \simeq  \, 
\frac{1+\rho}
{E_g \, m_b \, (1+2\rho)
\left[
1 \, - \, \sqrt{1  - \frac{m_g^2}{E_g^2}} \, \frac{ 1 - 2 t }{1+2\rho}
\right]} ,
\eea
where we have taken the strange quark after gluon emission to be on-shell, i.e. $p_s^2 = m_s^2$, 
since we are considering a {\it single primary} emission.
On the last member we have made the soft approximations for $E_g \ll E_s$ discussed above
and we have dropped the $m_g^2$ term, which comes out to be negligible 
{\it a posteriori} (see later).  
Since infrared singularities originate from quasi-real partons, i.e. from
partons with a small virtuality, we can assume that $m_g^2 \ll E_g^2$, 
to obtain \cite{landau4}:
\be
\Delta_s(p_s+p_g) \, \simeq \, 
\frac{ (1+\rho)^2 }{  m_b^2 \, \omega 
\left(
t \, + \, \rho
\right)
\left[ 
1 \, + \, \frac{ (1 + \rho)^2 \, m_g^2 }{ ( t \, + \, \rho ) \, \omega^2 \, m_b^2   }
\right] } \, ,
\ee
where we have made the small-angle approximation $1-2t \simeq 1$ in the term multiplying $m_g^2$,
since the latter is already a correction.
Now it comes the main point:
the $\mathcal{O}(\alpha_S)$ correction involves a real gluon, having $m_g^2 = 0$,
and is logarithmically enhanced in the infrared regions discussed in the previous 
section. In order not to spoil the lowest-order enhancement, the denominator has to remain small, 
i.e. the correction term in the square bracket has to be much smaller than one:
\be
\label{mgsmall}
m_g^2 \, < \, m_b^2 \, \omega^2  \, \frac{t + \rho}{(1+\rho)^2}  \, , 
\ee
where we have replaced the strong inequality with an ordinary inequality since we work
within logarithmic accuracy. 
The condition above specifies the upper limit on the gluon mass which does not
spoil the leading logarithmic structure of the cascade.
It is immediate at this point to justify neglecting the $m_g^2$ term on the
last member of eq.~(\ref{varieapprox}):
\be
\frac{m_g^2}{2 p_s \cdot p_g} \, \lsim \, \omega \, \ll \, 1 \ .
\ee
The amplitude for the emission of a soft gluon off the initial $b$ quark (at rest)
contains the scalar part of the $b$ propagator 
\be
\Delta_b( p_b - p_g ) \, = \, 
\frac{1}{ ( p_b - p_g )^2 - m_b^2 } 
\, = \, 
\frac{1}{ - \, 2 \, m_b \, E_g \, + \, m_g^2 } \, ,
\ee
which has a different enhancement with respect to the $s$ one; in particular, it
involves only the gluon energy and not the angle with respect to the $s$ quark.
However, for $m_g^2 \ne 0$, the kinematical constraint of fixed $y$ is no more
linear in $\omega$ (cfr. eq.~(\ref{expressy})) and reads:
\be
y \, \simeq \, \frac{\omega}{1+\rho} \, 
\left[ t + \rho  + \frac{ (1+\rho)^2 m_g^2 }{ m_b^2 \, \omega^2} \right] \, .
\ee
In particular, for $y \to 0^+$ one does not select anymore 
the limit $\omega \to 0^+$ unless the gluon mass satisfies the constraint
(\ref{mgsmall}).
As a consequence, for kinematical reasons, also amplitudes for soft emission
off the $b$ are enhanced only if condition (\ref{mgsmall}) holds.

The conclusion is that, as long as the restriction (\ref{mgsmall}) is satisfied,
one can sum over all possible gluon decay channels. 
The $y$-distribution is an example of {\it ``inclusive gluon decay''} quantity. 
By including {\it real} as well as {\it virtual} gluon-splitting processes,
the sum of cuts of the gluon propagator can be written as its discontinuity, 
so that the tree-level QCD coupling is replaced by the
following momentum-dependent effective coupling \cite{Amati:1980ch}:
\be
\label{replace}
\alpha_S \, \to \, \tilde{\alpha}_S\left( k^2 \right)
\, = \, \frac{i}{2\pi}
\int_0^{k^2} d m_g^2 \, {\rm Disc} 
\, \left[ \frac{1}{ m_g^2 \, \beta_0 \log \left( \frac{ - \, m_g^2}{ \Lambda_{QCD}^2 } \right) } \right]
\, \simeq \, \alpha_S(k^2),
\ee
where $\beta_0 = (11/3 C_A - 2/3 n_f)/(4\pi)$ with $C_A = N_c =  3$ and
we have defined the transverse momentum squared as the largest possible $m_g^2$:
\be
\label{defk2}
k^2 \, \equiv \, \max \, m_g^2 \, = \, m_b^2 \, \omega^2  \frac{t + \rho}{(1+\rho)^2} \, .
\ee
On the last member of eq.~(\ref{replace}), absorptive effects (the ``$-i\pi$'' terms in gluon polarization
function) have been neglected \cite{Aglietti:2004fz}
\footnote{The over-all sign on the r.h.s. of eq.~(11) of \cite{Aglietti:2004fz} 
should be minus.}.
Eq.~(\ref{replace}) implies that, in order to include leading gluon-decay effects, 
one has to evaluate the running coupling at the transverse momentum 
squared. The latter has the range: 
\be 
\frac{m_b^2 \, y^2}{1+\rho} \, < \, k^2 \, < \, \frac{m_b^2 \, y^2}{\rho} \, . 
\ee
Let us note that this range shrinks to zero for $\rho \to \infty$,
killing any collinear enhancement.
Finally, let us observe that in the massless limit for the strange quark, 
$\rho = 0$, one recovers the definition of transverse momentum given in 
\cite{catanitrentadue}.

\subsection{Effective Single-Gluon Distribution}

In this section we include gluon branching effects, i.e. secondary emission, 
in the single-gluon distribution by means of the replacement (\ref{replace}).
The real contribution to the form factor therefore reads:
\be
\label{nonintegr}
\sigma^{(R)}_S(y)
\, = \, 
\frac{C_F}{\pi} \, \int_{\rho}^{1+\rho} d\tau \, \int_0^1 \frac{d\omega}{\omega} 
\left[ 
 \frac{1 + 2\rho}{ \tau }
 - 1 - \, \frac{\rho(1+\rho)}{ \tau^2 }
\right] \, 
\alpha_S\left[ \frac{m_b^2 \, \omega^2 \, \tau}{(1+\rho)^2} \right]
\delta \left( y \, - \, \frac{\omega \, \tau}{1+\rho} \right) \, ,
\ee
where we have changed variable from $t$ to 
\be
\tau \, \equiv \, t \, + \, \rho \, .
\ee
Let us consider the integration of the above terms in turn:
\begin{enumerate}
\item
after integrating over $\tau$ and changing variable from $\omega$ to $k^2$, 
the first term in the square bracket on the r.h.s. of eq.~(\ref{nonintegr}) 
reads: 
\be
\label{dlsoft}
\frac{1}{y} \, \frac{C_F}{\pi} \, (1+2\rho) 
\int_{ \frac{m_b^2 \, y^2}{1+\rho} } ^{ \frac{ m_b^2 \, y^2 }{ \rho } } 
\frac{dk^2}{k^2} \, \alpha_S(k^2) \, .
\ee
Note that for $\rho \gg 1$ the coupling becomes large and
we leave the perturbative phase of QCD even outside the threshold region,
i.e. for large $y \sim \mathcal{O}(1)$;
\item
the second term on the square bracket (related to soft-gluon emission off the $b$), 
after integrating over $\omega$ the $\delta$-function, reads: 
\be
- \frac{1}{y}
\, \frac{C_F}{\pi}
\int_{\rho}^{ 1+\rho } d \tau \, \alpha_S \left( \frac{m_b^2 y^2}{ \tau } \right)
\, \cong \, 
- \, \frac{1}{y} \, \frac{C_F}{\pi} \, \alpha_S \left( \frac{m_b^2 \, y^2}{1+\rho} \right) \, ;
\ee 
\item
the last term in the square bracket (related to soft gluon emission off the {\it s} quark) 
is treated in a similar manner and reads:
\be 
- \frac{C_F}{\pi} \frac{ \rho(1+\rho) }{y} 
\int_{\rho}^{1+\rho}
\frac{d\tau}{\tau^2} 
\, \alpha_S\left( \frac{m_b^2 y^2}{ \tau } \right)
\, \cong \, 
- \, \frac{1}{y} \, \frac{C_F}{\pi} \, \alpha_S \left( \frac{ m_b^2 \, y^2 }{ \rho } \right) \, .
\ee
\end{enumerate}
Let us note that in 1. we have rewritten the logarithmic integration over $\tau$
in terms of $k^2$, while in 2. and in 3. we have explicitly computed the
non-logarithmic integrals over $\tau$. 

\noindent
By summing all three terms, we obtain:
\be
\sigma_S^{(R)}(y; \, \rho, m_b^2  ) \, = \,   
\frac{1}{y} 
\Bigg\{ 
\int_{ \frac{m_b^2 \, y^2}{ 1 + \rho } } ^{ \frac{ m_b^2 \, y^2 }{ \rho } } 
\frac{dk^2}{k^2} \, A^{(1)}(\rho) \, \alpha_S(k^2) 
\, + \,      D^{(1)} \alpha_S \left(  \frac{m_b^2 \, y^2}{ 1 + \rho } \, \right) 
\, + \, \Delta^{(1)} \alpha_S \left( \frac{ m_b^2 \, y^2 }{\rho} \right) 
\Bigg\} \, .
\ee
In the frozen coupling limit 
$\alpha_S \left( \cdots \!\,\right) \, \to \, \alpha_S$
or, equivalently, in first order, the
above expression reduces to the real contribution in the last member
 of eq.~(\ref{softff}): 
\be
\sigma_S^{(R)}(y, \rho, \, \alpha_S)
\, = \, 
\frac{C_F \alpha_S}{\pi} \,  \frac{1}{y} \, 
\left[
\big( 1 + 2 \rho \big) \, \log \left( \frac{1+\rho}{\rho} \right) \, - \, 2
\right] 
\, + \, \mathcal{O}\left( \alpha_S^2 \right) \, .
\ee

\subsection{$N$-space and Exponentiation}
\label{Nspace}

The exponentiation of threshold form factors in gauge theories
is a consequence both of dynamical and kinematical factorization
properties \cite{Bassetto:1984ik,Catani:1997vp,qcdgeneral}.
In QED, multiple soft-photon matrix elements factorize into the product
of single-photon matrix elements because of independent emission (Poisson distribution).
In QCD that is no longer true: soft gluons are not emitted independently because
of gluon splitting, but gluon correlations largely cancel in the form
factor, leading to an ``effective'' matrix element factorization.
On the kinematical side, to have exponentiation, the constraint for multiple soft emission
must factorize into the product of single-gluon (or single-photon) kinematical constraints.
The latter property does not hold in the physical, energy-momentum, space and, to have
a consistent resummation, one has to transform the form factor to $N$-space:
\be
\sigma_{S, \, N} \, \equiv \, \int_0^1 (1 - y)^{N-1} \, \sigma_S(y) \, dy .
\ee
The QCD form factor in the original (physical) space is recovered by
means of a standard inverse transform:
\be
\sigma_S(y) \, = \, \frac{1}{2\pi i} \int_{C-i\infty}^{C+i\infty} d N (1-y)^{-N} \, \sigma_{S, \, N} \, ,
\ee
where $C$ is a real constant chosen so that all the singularities of 
$\sigma_{S,N}$ lie to the
left of the integration contour.

Up to now we computed only real emission contributions to the form factor.
Virtual corrections do not need to be explicitly evaluated:
they can be included by imposing the vanishing of the first moment ($N = 1$)
of the form factor, implying the replacement 
\be
(1-y)^{N-1} \, \to \, ( 1 - y )^{N-1} \, - \, 1 \, .
\ee
By inserting the expression for $\sigma_S^{(R)}(y)$ obtained in the previous section
and exponentiating the result in order to take into account multiple primary emissions 
(real and virtual ones), we obtain:
\bea
\label{LLs}
\sigma_{S,N}(\rho,m_b^2) & = & \exp 
\int_0^1 \frac{dy}{y} \Big[ (1-y)^{N-1} - 1 \Big] 
\Bigg\{
\int_{ \frac{m_b^2 \, y^2}{1+\rho} }^{ \frac{ m_b^2 \, y^2 }{ \rho } } 
A^{(1)}(\rho) \, \alpha_S(k^2) \, \frac{dk^2}{k^2} \, +
\nn\\ 
&& ~~~~~~~~~~~~~~~~~~~~~~~~
+ \, {D}^{(1)} \, \alpha_S \left( \frac{m_b^2 \, y^2}{1+\rho} \right)
\, + \, \Delta^{(1)} \, \alpha_S \left( \frac{ m_b^2 \, y^2 }{\rho} \right)
\Bigg\} \, .
\eea
Eq.~(\ref{LLs}) consistently resums soft logarithms in leading
logarithmic ($LL_S$) approximation
\footnote{
The form of the resummation coefficients given in eq.~(\ref{LLs}) is not unique.
An alternative form involves a coefficient $\tilde{A}^{(1)}$ 
equal to the massless one and coefficients $\tilde{D}^{(1)}$ and $\tilde{\Delta}^{(1)}$ depending on
$\rho$:
\be
\tilde{A}^{(1)} \, = \, \frac{C_F}{\pi} \, ;
~~~~~
\tilde{D}^{(1)}(\rho) \, = \, - \, \frac{C_F}{\pi} \, \left( 1 - \rho \log \frac{1+\rho}{\rho}   \right) \, ;
~~~~~
\tilde{\Delta}^{(1)}(\rho) \, = \, - \, \frac{C_F}{\pi} \, \left( 1 - \rho \log \frac{1+\rho}{\rho}   \right) \, .
\ee
Note that $\tilde{D}^{(1)}(\rho)$ and $\tilde{\Delta}^{(1)}(\rho)$ vanish for $\rho \to \infty$.},
i.e. all the soft terms having the form
in $N$-space
\be
\alpha_S^n \log^n N \, . 
\ee
For a complete next-to-leading logarithmic ($NLL_S$) approximation, i.e.
the inclusion of all the terms 
\be
\alpha_S^n \log^{n-1} N \, ,
\ee
one has to include second-order corrections to the above first-order
terms, i.e.
\be
A^{(1)}(\rho)  \,\alpha_S  \to  A^{(1)}(\rho)  \,\alpha_S  +  
A^{(2)}(\rho)  \,\alpha_S^2 \,;\,\,\,~
%~~~~
D^{(1)}  \alpha_S  \to  D^{(1)}  \alpha_S  +  
D^{(2)}  \alpha_S^2 \,;\,\,\,~
%~~~~
\Delta^{(1)}  \alpha_S  \to  \Delta^{(1)}  \alpha_S  +  
\Delta^{(2)}  \alpha_S^2 \,.
\ee
These higher-order corrections describe residual multi-parton
interactions which are not taken into account neither by the exponentiation
of the one-gluon distribution nor by the insertion of the effective gluon-jet coupling
$\alpha_S(k^2)$
\footnote{
That is similar to the (more familiar) inclusion of higher-order corrections 
to a one-loop anomalous dimension:
$\gamma^{(0)} \, \alpha_S \to \gamma\left(\alpha_S\right) = 
\gamma^{(0)} \, \alpha_S + \gamma^{(1)} \, \alpha_S^2 + \cdots$.
}.
The second-order corrections are, as far as we know, not available at present in QCD 
\footnote{
Virtual corrections to heavy flavor decay have been computed in two-loop order
in soft approximation within an effective theory framework for $n_f = 0$ 
(purely gluonic case) in \cite{Korchemsky:1987wg}.
The relation to QCD distributions is straightforward to $\mathcal{O}(\alpha_S)$,
but beyond leading order we do not fully understand the regularization-renormalization scheme 
effects.}.
However, an improvement of eq.~(\ref{LLs}) can be obtained by using the two-loop approximation
for the QCD coupling $\alpha_S$ and by adding the second-order correction
$A^{(2)}(\rho)$ in the limit $\rho \to 0$, which is known since a long time 
\cite{Kodaira:1981nh,Kodaira:1982cr,Catani:1988vd}:
\be
\label{A2}
A^{(2)}(\rho) \, \approx \, A^{(2)}(0) \, = \, 
\frac{C_F}{\pi^2} \left[ C_A\left( \frac{67}{36} - \frac{z(2)}{2} \right)
- \frac{5}{18} n_f \right] \, ,
\ee
where $z(2) = \pi^2/6$ and $n_f$ is the number of active quark flavors.
Let us note that the term proportional to $A^{(2)}$ gives a contribution 
on the r.h.s. of eq.~(\ref{LLs})
\be
\approx \, A^{(2)} \, \alpha_S^2 \, \log \left( \frac{1+\rho}{\rho} \right) \, ,
\ee
which vanishes in the limit $\rho \to \infty$, and therefore does not spoil the vanishing of the exponent 
of the form factor in that limit (see eq.~(\ref{rto1})).
Soft logarithms can formally be resummed to
any accuracy by including higher order corrections to the above functions, 
namely:
\bea
\label{mainsec}
\sigma_{S,N}(\rho, m_b^2) & = & \exp 
\int_0^1 \frac{dy}{y} \Big[ (1-y)^{N-1} - 1 \Big] 
\Bigg\{
\int_{ \frac{m_b^2 \, y^2}{1+\rho} }^{ \frac{ m_b^2 \, y^2 }{\rho} } 
A \left[ \rho ; \, \alpha_S(k^2) \right] \, \frac{dk^2}{k^2} \, +
\nn\\
%&& ~~~~~~~~~~~~~~~~~~~
&\,+ \,& D \left[ \alpha_S \left( \frac{m_b^2 \, y^2}{1+\rho} \right) \right]
\, + \, \Delta \left[ \alpha_S \left( \frac{ m_b^2 \, y^2 }{\rho} \, \right) \right]
\Bigg\} \, .
\eea
The functions $D\left(\alpha_S \right)$ and $\Delta\left(\alpha_S \right)$
coincide up to first order and, on the basis of physical arguments,
it was conjectured in \cite{Aglietti:2006wh} that this property extends to higher orders, 
i.e. that
\footnote{
An explicit check of the above relation requires a massive two-loop computation,
which is outside the scope of the present paper.}
\be
\Delta\left(\alpha_S \right) \, = \, D\left(\alpha_S \right) \, . 
\ee
We do not deal with this problem here because $A^{(2)}(\rho)$ is not available
at present and therefore our analysis is limited to including 
$D^{(1)}$ and $\Delta^{(1)}$ only.

The soft resummation formula in eq.~(\ref{mainsec}) is the main result of this section.
Let us comment upon it.
The term proportional to $A(\rho;\alpha_S)$ contains a logarithmic integration
over the gluon transverse momentum squared and becomes large for $\rho$ small:
it contains a kind of ``$\rho$-evolution'', while in the massless case one
has an evolution in $y$ (cfr. eq.~(\ref{citoprima})).
The contribution in $D(\alpha_S)$ comes from soft emission off the beauty quark,
while that one in $\Delta(\alpha_S)$ from the strange quark.

\section{Resumming Final-Mass Effects}
\label{sec3}

In this section we derive a general resummation formula 
by using the dipole factorization formulae for the massive case
derived in \cite{Dittmaier:1999mb,Catani:2002hc} (see also \cite{Keller:1998tf}).
By taking the photon in the final state as the spectator parton, one obtains
from eq.~(5.16) of \cite{Catani:2002hc}:
\bea
\label{dipoli}
\sigma^{(R)}(y)
&=& 
\frac{1}{y} \, \frac{C_F  \, \alpha_S}{\pi} \, \int_{ \frac{\rho}{ y + \rho } }^1 
\left[
\,  \frac{ 1 + 2\rho }{ 1 - (1 - y) \, z } 
\, - \, \frac{1 + z}{2} 
\, - \, \frac{\rho}{y}
\, \right] \, dz \, +
\nn\\
&& ~~
+ ~ {\rm (soft~emission~contributions~off~the~} b, \, \propto \, D^{(1)}),
\eea
where $z$ is the normalized projection of the strange 4-momentum along the (reference) 
photon 4-momentum:
\be
z \, = \, z_s \, \equiv \, \frac{ p_s \cdot p_{\gamma} }{ (p_s + p_g) \cdot p_{\gamma} } \, ;
~~~~~~~
z_g \, \equiv \, \frac{ p_g \cdot p_{\gamma} }{ (p_s + p_g) \cdot p_{\gamma} } \, = \, 1 - z_s \, .
\ee
The factor $1+2\rho$ is not present in eq.~(5.16) of \cite{Catani:2002hc}: 
we have inserted it by hand in order to take into account the specific {\it soft} structure 
of heavy flavor decay, namely one heavy leg in the initial state and one lighter 
leg in the final state; this process was not explicitly considered in \cite{Catani:2002hc}.
\footnote{
Let us stress that $1+2\rho \to 1$
in the quasi-collinear limit ($k_{\perp}\to 0, \, m \to 0$ with $k_{\perp}/m \to$ const).
That implies that this factor cannot be determined with quasi-collinear 
factorization but only with soft factorization because it is process dependent.
}
Let us now sketch the integrations of the above terms in turn:
\begin{enumerate}
\item
by keeping logarithmic terms only, the first term in the square bracket on the r.h.s. 
of eq.~(\ref{dipoli}) reads:
\be
\label{dl}
\frac{C_F \alpha_S}{\pi} \, \frac{1+2\rho}{y} \, \int_y^{ \frac{(1+\rho) y}{y+\rho} } 
\frac{d\omega}{\omega}
\, = 
\, - \, \frac{C_F \alpha_S}{\pi} \, \frac{1 + 2\rho}{y} \, \log \left( \frac{y+\rho}{1+\rho} \right) \, ,
\ee
where the unitary gluon energy reads in terms of light-cone variables:
\be
\label{relate}
\omega \, \equiv \, 1 \, - \, ( 1 - y ) \, z \, \simeq \, 1 \, - \, z \, + \, y \, .
\ee
Eq.~(\ref{relate}) is easily proved by using the following expression for $y$:
\be
y \, = \, y_{s g ,\gamma} \, = \, 
\frac{p_s\cdot p_g}{ p_s \cdot p_{\gamma} \, + \, p_g\cdot p_{\gamma} \, + \, p_s\cdot p_g} \, ;
\ee
\item
The second term on the r.h.s., related to hard collinear emission off the $s$ quark, reads:
\be
- \, \frac{C_F\alpha_S}{\pi}
\left[ \,
\frac{3}{4} \frac{1}{ y + \rho }
\, + \, \frac{ \rho }{ 4 } \, \frac{1}{ \left( y + \rho \right)^2 }
\, \right] \, .
\ee  
The second term in the above equation is always smaller than the first one and it 
does not produce any logarithmic contributions. 
Therefore, it can safely be neglected.
For $y \, \gg \, \rho$, we re-obtain the well-known massless contribution, 
$ - 3 C_F \alpha_S/(4 \pi \, y)$, while in the opposite regime,
$ y \, \ll \, \rho$,
the above contribution can be neglected because there is no more a small denominator.
Let us stress that all that is in complete agreement with physical intuition;
\item
The last term on the r.h.s., related to soft emission not collinearly enhanced
off the strange quark, reads:
\be
- \frac{C_F \alpha_S}{\pi} 
\left( 
\frac{1}{y} - \frac{1}{ y + \rho } 
\right) \, .
\ee
\end{enumerate}

\subsection{Inclusive Gluon-Decay Effects}

The gluon transverse momentum squared in conveniently written as
\be
\label{newk2}
k^2 \, \equiv \, \frac{m_b^2 \, y \, \omega}{1+\rho} \, ,
\ee
where $\omega$ is expressed in terms of light-cone variables in eq.~(\ref{relate})
(see also \cite{Gieseke:2003rz}).
As in the soft case, we include gluon-branching effects by replacing the
tree-level coupling with the effective coupling (\ref{replace}) in the one-gluon
distribution (\ref{dipoli}):
\begin{enumerate}
\item
The double-logarithmic term on the l.h.s. of eq.~(\ref{dl}) then reads:
\be
\frac{1}{y} \, \frac{C_F}{\pi} \, (1+2\rho) \, 
\int_{ \frac{m_b^2 y^2}{1+\rho} }^{ \frac{ m_b^2 y^2 }{ y \, + \, \rho } } 
\frac{dk^2}{k^2} \, \alpha_S(k^2) \, .
\ee
In the soft limit ($y \to 0$ at fixed $\rho\in(0,\infty)$), the above expression reduces to the 
corresponding eikonal contribution derived in eq.~(\ref{dlsoft}) of the previous section:
the only difference lies in the upper limit.
In the massless limit ($\rho \to 0$ with $y \ne 0$), the above formula reduces instead
to the standard (massless) contribution:
\be
\label{citoprima}
\frac{1}{y} 
\int_{m_b^2 y^2 }^{ m_b^2 y } \frac{dk^2}{k^2} \, A^{(1)} \, \alpha_S(k^2) \, ;
\ee
\item
The determination of the argument of the QCD coupling entering the subleading
terms goes as follows:
\bea
\label{3membri}
\hskip -1.2truecm
&& - \, \frac{1}{y} \, \frac{C_F  }{\pi} \, \int_{ \frac{\rho}{ y + \rho } }^1 
\left(
\, \frac{1 + z}{2} 
\, + \, \frac{\rho}{y}
\, \right) 
\, \alpha_S 
\left[
\frac{m_b^2 \, y \, \big( 1 - (1 - y) z \big)}{1+\rho}
\right]
\, dz 
\nn\\
&=& 
- \, \frac{1}{y} \, \frac{C_F  }{\pi} \, 
\, \alpha_S 
\left(
\frac{ m_b^2 \, y^2 }{ y \, + \, \rho }
\right)
\int_{ \frac{\rho}{ y + \rho } }^1 
\left(
\, \frac{1 + z}{2} 
\, + \,  \frac{\rho}{y}
\, \right) \, dz
\, + \, \mathcal{O} \left( \frac{\alpha_S^2}{y}  \right) \, ,
\eea
where on the last member we have replaced the following expansion of
the QCD coupling:
\be
\label{aSexpand}
\alpha_S \!\!
\left[
\frac{m_b^2  y \big( 1 - ( 1 - y ) z \big)}{1+\rho}
\right]\!
= 
\alpha_S\! 
\left(
\frac{ m_b^2  y^2 }{  y + \rho }
\right) 
- \beta_0
\alpha_S^2\! 
\left( \frac{ m_b^2 y^2 }{  y + \rho } 
\right) 
\log \!
\left\{\! \frac{ ( y + \rho ) \big[ 1 - ( 1 - y ) \, z \big]  }{ (1+\rho) y } 
\!\right\}
+ \cdots .
\ee
The $z$-integral on the r.h.s. of eq.~(\ref{3membri}) is equal to the one in lowest
order we have already calculated and the higher-order terms in the expansion (\ref{aSexpand})
produce terms beyond our accuracy, namely $\mathcal{O} ( \alpha_S^2/y )$.
Let us remark that the rule (\ref{replace}) can be applied also to describe the branching of a
hard gluon emitted collinearly to the $s$-quark, because the enhancement is given again
by the denominator of the $s$-quark propagator (see sec.~\ref{igd}).
\end{enumerate}
By explicitly adding the soft emission contribution off the $b$ obtained in the
previous section, the complete distribution reads:
\bea
\sigma^{(R)} (y; \rho, m_b^2) 
&=& 
\frac{1}{y} \, 
\int_{ \frac{m_b^2 \, y^2}{1+\rho} }^{ \frac{ m_b^2 \, y^2 }{ y \, + \, \rho } } 
A^{(1)} \left( \rho \right) \, \alpha_S(k^2) \, \frac{d k^2}{k^2} \, + \, 
D^{(1)} \, \alpha_S \left( \frac{m_b^2 \, y^2}{1 + \rho} \right) \, \frac{1}{y} \, +
\\
&+& \Delta^{(1)} \alpha_S\left( \frac{ m_b^2 y^2 }{y + \rho } \right)  
\left( \frac{1}{y} - \frac{1}{ y + \rho } \right) 
+ B^{(1)} \alpha_S \left( \frac{ m_b^2 y^2 }{ y + \rho } \right) 
\frac{1}{ y + \rho } .
\nn
\eea

\subsection{$N$-Space and Exponentiation}

By repeating the computation made in the soft case, one obtains:
\bea
\hskip -1truecm
\sigma_N (\rho ; \, m_b^2) \!\!&=&\!\! \exp \int_0^1 dy \big[ (1-y)^{N-1} - 1 \big]
\Bigg\{
\frac{1}{y}
\int_{ \frac{m_b^2 y^2}{1+\rho} }^{ \frac{ m_b^2 \, y^2 }{ y + \rho } } 
A^{(1)} \left( \rho \right) \alpha_S(k^2) \frac{d k^2}{k^2} +
\frac{1}{y} D^{(1)} \alpha_S\left( \frac{m_b^2 y^2}{1+\rho} \right) +  
\nn\\
&&~~~~~~~~~~  + \, \left( \frac{1}{y} \, - \, \frac{1}{ y \, + \, \rho } \right) \,
\Delta^{(1)} \alpha_S\left( \frac{ m_b^2 \, y^2 }{y + \rho } \right)  \, + \,
\frac{1}{ y \, + \, \rho } \, B^{(1)} 
\, \alpha_S \left( \frac{ m_b^2 \, y^2 }{ y + \rho } \right)   
\Bigg\} \, .
\eea
What about the inclusion of higher orders?
The first point is that there is a different counting in the soft region $\rho \gsim \mathcal{O}(1)$,
which is a single logarithmic region, and in the collinear ($\rho=0$) or quasi-collinear regions ($\rho \ll 1$), 
which are double logarithmic ones.
For instance, for large $\rho \approx \mathcal{O}(1)$, the term proportional to $A^{(1)}(\rho)$ is of the same order
of the $D^{(1)}$ and $\Delta^{(1)}$ terms. On the contrary, for $\rho \ll 1$ the $A^{(1)}$ term is leading
because it is multiplied by a double log, while $D^{(1)}$ and $\Delta^{(1)}$ are 
subleading because they are multiplied by a single log. 
That complication arises because our formula is a smooth interpolation between two different
dynamical regions.
An improvement can be realized as in the soft case by means of the replacement:
\be
A^{(1)}(\rho) \, \alpha_S \, \to \,  A^{(1)}(\rho) \, \alpha_S \, + \, A^{(2)}(0) \, \alpha_S^2\, ,
\ee
with $A^{(2)}(0)$ given in eq.~(\ref{A2}).
The above substitution allows a {\it complete} $NLL$ resummation in the {\it massless} case
($\rho/y \to 0$) and a {\it partial} $NLL$ resummation in the {\it soft} case ($y/\rho \to 0$).

Infrared logarithms of subleading orders can be formally resummed by including higher-order
corrections to the functions in the exponent, similarly to the soft case:
\bea
\label{quasifinale}
\hskip -1truecm
\sigma_N(\rho ; m_b^2)
\!\!&=&\!\! \exp \int_0^1 \!dy \, \Big[ (1-y)^{N-1} - 1 \Big]
\Bigg\{
\frac{1}{y} 
\int_{ \frac{m_b^2 y^2}{1+\rho} }^{ \frac{ m_b^2 \, y^2 }{ y \, + \, \rho } } 
\!A\left[ \rho ; \, \alpha_S(k^2) \right]  \frac{d k^2}{k^2} 
+ \frac{1}{y} D\left[ \alpha_S \left( \frac{m_b^2 \, y^2}{1+\rho} \right)\right] \, +
\nn\\
&& ~~~~~~~~~~~  + \, \left( \frac{1}{y} \, - \, \frac{1}{ y \, + \, \rho } \right)
\Delta \left[ \alpha_S \left( \frac{ m_b^2 \, y^2 }{ y + \rho } \right) \right]
\, + \, \frac{1}{ y \, + \, \rho } \,
B \left[ \alpha_S \left( \frac{ m_b^2 \, y^2 }{ y + \rho } \right) \right] \,
\Bigg\}\, .
\eea

\subsection{A Symmetry of the Resummation Functions}
\label{secinv}

Eq.~(\ref{quasifinale}) is invariant under the simultaneous substitutions:
\bea
A(\rho; \alpha_S) &\to& A(\rho,\alpha_S) \, + \, \beta(\alpha_S) \, \frac{d}{d\alpha_S} F\left(\alpha_S\right) \, ;
~~~~~
D(\alpha_S) ~ \to ~  D(\alpha_S) \, + \, F(\alpha_S) \, ;
\nn\\
\Delta(\alpha_S) &\to& \Delta(\alpha_S) \, - \, F(\alpha_S) \, ;
~~~~~~~~~~~~~~~~~~~~~~ \, 
B(\alpha_S) ~ \to ~  B(\alpha_S) \, - \, F(\alpha_S) \, ,
\eea
where $F(\alpha_S)$ is an arbitrary function having a power-series expansion in the
QCD coupling,
\be
F(\alpha_S) \, = \, \sum_{n=1}^{\infty} F^{(n)} \, \alpha_S^n \, .
\ee
We have defined the QCD $\beta$-function coefficients with an over-all minus sign:
\be
k^2 \frac{d\alpha_S}{d k^2} \, = \, \beta(\alpha_S) 
\, = \, - \, \beta_0 \, \alpha_S^2 \,  - \, \beta_1 \, \alpha_S^3 \, - \, \cdots \, .
\ee
The above relation implies that only 3 out of the 4 functions entering eq.~(\ref{quasifinale})
are independent; in particular, one can choose $F(\alpha_S)$ in such a way as to make one out of the
functions $B(\alpha_S)$, $D(\alpha_S)$ and $\Delta(\alpha_S)$ vanishing, or to reduce 
the function $A(\rho,\alpha_S)$ to its lowest-order term $A^{(1)}(\rho)\, \alpha_S$.
The above symmetry is a generalization of that one of the massless formula ($\rho=0$)
involving only the functions $A$, $B$ and $D$ found in \cite{Catani:1990rp}.
In NLL, the above substitutions read:
\be
A^{(2)}(\rho) \, \to \,  A^{(2)}(\rho) \, - \, \beta_0 \, F^{(1)} \, ;
~~~
D^{(1)} \, \to \,  D^{(1)} \, + \, F^{(1)} \, ;
~~~
\Delta^{(1)} \, \to \,  \Delta^{(1)} \, - \, F^{(1)} \, ;
~~~
B^{(1)} \, \to \,  B^{(1)} \, - \, F^{(1)}  .\!
\ee
Finally, the resummation coefficients are also modified by a change of the
renormalization scheme of $\alpha_S = {\alpha_S\,}'(1 + C {\alpha_S\,}' + \cdots)$; 
this problem has been investigated in 
\cite{Catani:1990rr}.

\section{Semileptonic Decay}
\label{sec4}

This is the central section of the paper, in which we extend
threshold resummation from the radiative decay to the
semileptonic one.
By means of general field-theoretic arguments, one can show that the hard scale $Q$ in
heavy flavor decays is proportional to the total final hadron energy 
\cite{Aglietti:2001br,Aglietti:2005mb}:
\be
\label{vecchia}
Q \, = \, 2 \, E_{X_f} \, + \, \mathcal{O} \left( \frac{m_{X_f}^2}{E_{X_f}^2}  \right) \, ,
\ee
where $f = s, c$.
The main difference between the radiative decay ($q^2 = 0$, $q = p_{\gamma}$) and the semileptonic one 
($q^2 \ge 0$, $q = p_l + p_{\nu}$)
is that in the latter case the lepton-neutrino pair can take away from the QCD final state 
a large fraction of the initial energy provided by the $b$-quark mass.
We may fix the power-suppressed terms in eq.~(\ref{vecchia}) by imposing that the hard scale
$Q$ exactly coincides with the beauty mass $m_b$ when $q^2 = 0$ 
(see eq.~(\ref{suggerimento})), so that: 
\be
Q \, = \, E_{X_f} \, + \, \left| \vec{p}_{X_f} \right| ~ \Rightarrow ~ m_b ~~~~ {\rm when} ~~ q^2 \, = \, 0 \, . 
\ee
The hadronic variable $y$ and the mass-correction parameter $\rho$ must therefore be defined 
by replacing $m_b$ with the general ($q^2$-dependent) hard scale $Q$: 
\be
y \, = \, \frac{m_{X_s}^2 - m_s^2}{ m_b^2 \, - \, m_s^2 } 
~~~ \to ~~~ \frac{m_{X_c}^2 - m_c^2}{ Q^2 \, - \, m_c^2 } \, ;
~~~~~~~~
\rho \, = \, \frac{m_s^2}{ m_b^2 \, - \, m_s^2} 
~~~ \to ~~~ \frac{m_c^2}{ Q^2 \, - \, m_c^2} \, .
\ee
One then obtains the resummed QCD form factor in $N$-space 
given in eq.~(\ref{main}) of the introduction, the main result of this work.
This equation systematically resums all threshold logarithms occurring in the semileptonic
$B$ decays for any value of the ratio $m_c/m_b \in [0,1]$.
It interpolates in a smooth way dynamical regions 1., 2. and 3. described in the introduction.

If the charm quark is not (too) fast, as it happens in most of the phase space of
semi-leptonic $b \to c$ transitions, one can use the simplified
resummation formula (\ref{simplified}), which resums soft logarithms only.
The latter correctly describes regions 1. and 2. but, unlike
the full result (\ref{main}), cannot account for the massless limit $\rho \to \infty$, 
i.e. for region 3.
In the soft limit:
\be
E_{X_c} \, \simeq \, E_c \, \simeq \, E_c^{(0)} \, = \, \frac{m_b^2+m_c^2-q^2}{2m_b} \, ;
~~~~~~~
\left| \vec{p}_{X_c} \right| \, \simeq \, \left| \vec{p}_{c} \right| \, 
\simeq \,\left| \vec{p}_c^{\,\,(0)} \right|  
\, = \, \frac{ \sqrt{ \lambda(m_b^2, m_c^2, q^2) } }{2m_b} \, ,
\ee
where $\lambda(a,b,c) \, \equiv \, a^2 + b^2 + c^2 - 2a b - 2 a c - 2 b c$.
It is straightforward to check that the above formulae reduce to the one for the
radiative case in the limit $q^2 \to 0 $.

\section{Conclusions}
\label{sec5}

We have presented in eq.~(\ref{finale}) an expression for the QCD form factor
resumming threshold logarithms in the semileptonic decay
\be
B \, \to \, X_c \, + \, l \, + \, \nu_l \, .
\ee
The resummation is valid for any chosen value of the ratio
$m_c/m_b \in [0,1]$ and reduces to the standard next-to-leading
logarithmic approximation for $m_c \to 0$.
Coherence effects, occurring for a small relative velocity of the $c$ and $b$
quarks and leading to destructive soft-gluon interference, 
are correctly incorporated in the form factor, whose corrections vanish in the no-recoil point.
Our form factor is basically a smooth interpolation of a {\it soft} resummation and a 
{\it soft + collinear} one and constitutes a consistent description of a multi-scale 
process like semileptonic $b \to c$ decay; such interpolation is not unique
and has been constructed on the basis of simplicity's requirements.
The resummation formula in eq.~(\ref{finale}) has been derived form ``first principles'',
namely from the universal properties of QCD radiation.
Much attention has been paid to the correct inclusion of gluon-branching effects
by means of a proper definition of gluon transverse momentum.
We have also presented in eq.~(\ref{simplified}) a simplified resummation formula which takes
into account soft effects only and which can be applied as long as the charm quark is not 
too fast, as is the case in most of the phase space of semileptonic $b\to c$ transitions.

Resummed perturbation theory breaks down for final hadron masses as small as
\be
\label{nopert}
m_X^2 \Big|_{NP} \, \approx \, m_c^2 \, + \, \Lambda_{QCD} \sqrt{Q^2 - m_c^2} \, , 
\ee
because of large soft effects controlled by the QCD coupling evaluated close to the 
Landau pole.
Such effects can formally be factorized by the shape function for a massive quark in the
final state \cite{Mannel:1994pm}.
In practice, one can model them with an effective ghost-less QCD coupling 
constructed on the basis of some analyticity requirement, along the lines of 
\cite{Aglietti:2006yb,Aglietti:2006yf}.
Our resummed calculation can be combined with the $\mathcal{O}(\alpha_S)$ triple differential distribution 
computed in \cite{Trott:2004xc,agru} by a proper matching procedure in order to obtain an uniform 
approximation in the whole phase space.

We shortly mention here some phenomenological outputs of our results.
Resummation effects are expected to be largely sensitive to the charm mass and 
a phenomenological analysis could probably results in a more precise determination of $m_c$. 
Unlike the charmless channel, the $b \to c$ rate is known to be dominated by few hadronic 
states; given the high-quality data of the $B$-factories,
comparison with solid theoretical expectations could also provide a non-trivial check
of local parton-hadron duality and, in general, it could give more information about hadron dynamics at the scale of few GeV's.
The theoretical knowledge of $b \to c$ semileptonic spectra could also be of utility for the inclusive extraction of the CKM 
matrix element $|V_{cb}|$.
Furthermore in~\cite{Aglietti:2006yb}  a model was formulated
which is in disagreement with the charmless electron spectra 
below $2.2~\rm{GeV}$ measured at the $B$-factories; 
a natural explanation is an under-subtracted charm background, 
which could be better estimated by means of our results.

Our formula can also be used to describe beauty-mass effects in semileptonic
top decays $t \to b +W$.
A natural extension of our formalism would also allow a smooth description of heavy-flavor production
in $e^+e^-$ annihilation,
\be
\label{HFP}
e^+ e^- \, \to \, Q \, \bar{Q} \, ,
\ee
as function of the center-of-mass energy $E_{c.m.}$, from threshold, $E_{c.m.} = 2m_Q$, up to 
higher energies, $E_{c.m.} \gg 2m_Q$.
As far as soft effects are concerned this process may be seen as a crossed semileptonic decay.
Coherence effects play a similar role as in the heavy flavor decay also in the production process (\ref{HFP}): 
one produces a $Q\bar{Q}$ pair in a color singlet state,
which at the threshold, classically, sitting one on the top of the other, do screen their color and consequently
emit no soft radiation.

\vskip 0.5truecm

\centerline{\bf Acknowledgments}

\noindent
We would like to thank V. Del Duca for discussions.

\noindent
G.~F. acknowledges support by the European Community's Marie-Curie
Research Training Network Programme under contract MRTN-CT-2006-035505
{\it ``Tools and Precision Calculations for Physics Discoveries at Colliders''.}

\end{document}